\documentclass[aps,prb,twocolumn]{revtex4}
\usepackage{graphicx}
\usepackage{natbib}

\begin{document}

\title{Temperature-dependent mechanisms for the dynamics of protein-hydration waters: \\ a molecular dynamics simulation study}

\author{M.\ Vogel}
\affiliation{Institut f\"ur Festk\"orperphysik, Technische Universit\"at Darmstadt,
Hochschulstr.\ 6, 64289 Darmstadt, Germany}

\begin{abstract}
Molecular dynamics simulations are performed to study the temperature-dependent dynamics and structures of the hydration shells of elastin-like and collagen-like peptides. For both model peptides, it is consistently observed that, upon cooling, the mechanisms for water dynamics continuously change from small-step diffusive motion to large-step jump motion, the temperature dependence of water dynamics shows a weak crossover from fragile behavior to strong behavior, and the order of the hydrogen-bond network increases. The temperature of the weak crossover from fragile to strong behavior is found to coincide with the temperature at which maximum possible order of the hydrogen-bond network is reached so that the structure becomes temperature independent. In the strong regime, the temperature dependence of water translation and rotational dynamics is characterized by an activation energy of $E_a\!\approx\!0.43\mathrm{\,eV}$, consistent with results from previous dielectric spectroscopy (DS) and nuclear magnetic resonance (NMR) studies on protein hydration waters. At these temperatures, a distorted $\pi$-flip motion about the twofold molecular symmetry axes, i.e., a water-specific $\beta$ process is an important aspect of water dynamics, at least at the water-peptide interfaces. In addition, it is shown that the hydration waters exhibit pronounced dynamical heterogeneities, which can be traced back to a strong slowdown of water motion in the immediate vicinity of peptide molecules due to formation of water-peptide hydrogen bonds. 

\end{abstract}

\date{\today}

\maketitle

\section{Introduction}

A number of anomalies distinguish water from other liquids.\cite{Kumar_JPCM_08,Mishima_NAT_98} For example, it was argued that supercooled water exhibits a fragile-to-strong transition (FST) in the vicinity of $T_x\!=\!228\mathrm{\,K}$.\cite{Angell_SCI_08,Ito_NAT_99} Here, FST means that supercooled water behaves as a fragile liquid above $T_x$, i.e., the temperature dependence of viscosity deviates from an Arrhenius law, while water is a strong liquid below $T_x$ and, hence, an Arrhenius law is obeyed.\cite{Angell_SCI_95} It was put forward that the proposed FST is related to a hypothesized liquid-liquid critical point at elevated pressure, which terminates a phase transition line between a low-density liquid (LDL) and a high-density liquid (HDL) at low and high temperatures, respectively.\cite{Xu_PNAS_05} However, direct observation of the proposed FST is not possible for bulk water because of inevitable crystallization.\cite{Mishima_NAT_98}

By contrast, crystallization can be suppressed in confinement so that the dynamics of supercooled waters are accessible down to the glass transition. Confined and interfacial waters are of enormous importance for biological, geological, and technological processes. While some experimental studies on such waters observed crossovers in the temperature dependence of a correlation time $\tau$ at $T\!\approx\!224\mathrm{\,K}$\cite{Liu_PRL_05,Chen_PNAS_06} or $T\!\approx\!200\mathrm{\,K}$,\cite{Bergman_NAT_00,Swenson_PRL_06} others provided evidence for an absence of any crossover.\cite{Sokolov_PRL_08} Several workers took the crossover at ca.\ $\mathrm{224\,K}$ as indication for a FST and related it to a liquid-liquid transition of confined waters.\cite{Liu_PRL_05,Chen_PNAS_06,Zanotti_EPL_05} Other workers challenged this conclusion and argued that the Arrhenius processes found below the crossovers are secondary ($\beta$) relaxations,\cite{Swenson_PRL_06,Sokolov_PRL_08} such as the Johari-Goldstein (JG) $\beta$ process,\cite{Capaccioli_JPCB_07} rather than the structural ($\alpha$) relaxation, which is difficult to observe at these temperatures.\cite{Cerveny_PRL_04}

Protein hydration waters are prominent examples of interfacial waters. Due to an interplay of protein and water dynamics, existence of a hydration shell is essential for the biological functions of proteins.\cite{Rupley_APC_91} Despite appreciable progress in recent years, the nature of this coupling is still controversially discussed. It was argued that the protein dynamics are "slaved" by the water dynamics.\cite{Fenimore_PNAS_04,Fenimore_PNAS_02} However, it is not clear which relaxation processes of protein (local or global) and water ($\alpha$ or $\beta$) are related.\cite{Fenimore_PNAS_04,Swenson_PRL_06,Doster_BBA_05,Doster_EBPJ_08,Ngai_JPCB_08,Khodadadi_JPCB_08} Valuable insights have been gained by characterizing the temperature dependence of protein and water dynamics, in particular during the so-called dynamical transition of proteins at ca.\ $200\mathrm{\,K}$,\cite{Doster_EBPJ_08,Bagchi_CR_05} which was reported to be associated with a freezing of biological functions.\cite{Rasmussen_NAT_92} Interestingly, it was postulated that the dynamical transition of proteins is related to a FST of the hydration waters.\cite{Chen_PNAS_06,Zanotti_EPJ_07} 

A fundamental understanding of the coupling of protein and water dynamics requires, first, assignment of the various relaxation processes to protein and water, respectively, and, second, identification of the processes as $\alpha$ or $\beta$ relaxations. Recently, we demonstrated that $^2$H NMR is a powerful tool for this purpose.\cite{Vogel_PRL_08} For elastin and collagen, two important proteins of the connective tissue, we found that, when the temperature is decreased, the mechanism for the reorientation of the hydration waters changes from an isotropic motion to an anisotropic motion, which is dominated by large-angle rotational jumps. The mechanism for the low-temperature motion is consistent with neither the $\alpha$ process nor the JG $\beta$ process of molecular glass-forming liquids,\cite{Roessler,Vogel_JCP_01} implying the existence of a water-specific $\beta$ process. 

Providing access to the complete microscopic information, molecular dynamics (MD) simulations enable straightforward assignment and identification of the various relaxation processes. A huge amount of simulation works focused on the behaviors of hydrated proteins under physiological conditions,\cite{Bagchi_CR_05} e.g., the mechanisms for hydrogen-bond (HB) breaking were shown to differ for bulk and hydration waters.\cite{Jana_JPCB_08} Computational studies on the temperature-dependent properties of hydrated proteins are more rare.\cite{Tarek_PRL_02,Tarek_EBPJ_08,Tournier_PRL_03,Tournier_BPJ_03,Smith_PTRS_04,Siebert_PRL_08,Oleinikova_PRL_05,Jana_JPCB_08} It was found that the protein dynamical transition is triggered by the relaxation of the protein-water HB network, which in turn is driven by water translational diffusion.\cite{Tarek_PRL_02,Tarek_EBPJ_08,Tournier_BPJ_03,Smith_PTRS_04} Moreover, the protein dynamic transition was found to occur at the temperature of both a dynamic crossover in the diffusivity of hydration water and a maximum of specific heat, suggesting a relation to the hypothesized liquid-liquid critical point of water and to an evolution of water from HDL to LDL.\cite{Kumar_PRL_06,Lagi_JPCB_08} These results imply that water plays an important role for protein dynamics. 

We use MD simulations to investigate the temperature-dependent interplay of protein and water dynamics for hydrated elastin and hydrated collagen. In doing so, we exploit that peptides (VPGVG)$_n$ have proven successful models of ideal elastin,\cite{Li_JACS_01,Schreiner_PRL_04,Baer_JPCB_06} while triple-helical molecules [3(POG)$_{n}$]\cite{Hyp} model ideal collagen.\cite{Mooney_BP_01,Mooney_BP_02,Bronco_JPCB_04}  The choice of the proteins elastin and collagen is motivated by both their biological importance and their use in our previous $^2$H NMR study\cite{Vogel_PRL_08} so that we expect synergetic effects from a combined experimental and theoretical approach. In the present contribution, we focus on the temperature-dependent translational and rotational motion of the hydration waters to improve our knowledge about the dynamics of supercooled bulk water and the role of water for the biological functions of proteins. The protein dynamics and the interplay of the components will be investigated in future work.  

\section{Methods}

We study three models with the same hydration level $h\!=\!0.3$ (g water/$\mathrm{1\,g}$ peptide). The smaller model of hydrated elastin (E1) is comprised of 1 peptide (VPGVG)$_{50}$ and 342 H$_2$O, while the larger model (E8) contains 8 peptides (VPGVG)$_{50}$ and 2732 H$_2$O. The model of hydrated collagen (C5) is comprised of 5 triple-helical molecules [3(POG)$_{10}$] and 705 H$_2$O.

MD simulations were performed using the GROMACS software package.\cite{GROM} The GROMOS96 43a2 force field was utilized.\cite{GRO96} Moreover, the SPC model of water was employed,\cite{SPC} as recommended for GROMOS force fields. We applied periodic boundary conditions and a time step of $2\mathrm{\,fs}$. The nonbonded interactions were calculated utilizing a cutoff distance of $1.2\mathrm{\,nm}$. The PME technique\cite{PME} was used to treat the Coulombic interactions and the LINCS\cite{LINCS} and SETTLE\cite{SETTLE} algorithms were applied to constrain the bonds of peptide and water, respectively. Prior to data acquisition, the systems were equilibrated in simulations at constant $N$, $P$, and $T$, using the Rahman-Parrinello barostat\cite{RP} and the Nos\'{e}-Hoover thermostat.\cite{NH} These equilibration runs, which spanned up to $100\mathrm{\,ns}$ for E1 and up to $40\mathrm{\,ns}$ for E8, allowed us to adjust the densities $\rho(T)$, which increase linearly by 5-7\% upon cooling in the studied temperature ranges. The sizes of the cubic simulation boxes amounted to about $3.3\mathrm{\,nm}$ and $6.7\mathrm{\,nm}$ for E1 and E8, respectively. In the case of C5, the atoms of the triple helices [3(POG)$_{10}$] were restrained to their positions in the crystal structure, suppressing diffusive motion of the peptides. To model a collagen microfibril, five parallel triple-helical molecules were pentagonally grouped, where the distance between the centers of adjacent molecules was ca.\ $\mathrm{1\,nm}$. The temperature-dependent spatial distribution of the water molecules about the triple helices was adjusted in equilibration runs at constant $N$, $P$, and $T$, which spanned up to $40\mathrm{\,ns}$. For C5, the sizes of the simulation box were $L_x\!\approx\!8.4\mathrm{\,nm}$ and $L_y\!\approx L_z\! \approx\!3.4\mathrm{\,nm}$. For all systems, the subsequent production runs were performed in the canonical ensemble, i.e., at constant $N$, $V$, and $T$, employing the Nos\'{e}-Hoover thermostat. The trajectories were saved every $1\mathrm{\,ps}$ for later analysis.

\section{Results}

\subsection{Structure of hydration waters}

\begin{figure}
\begin{center}
\includegraphics[width=7cm]{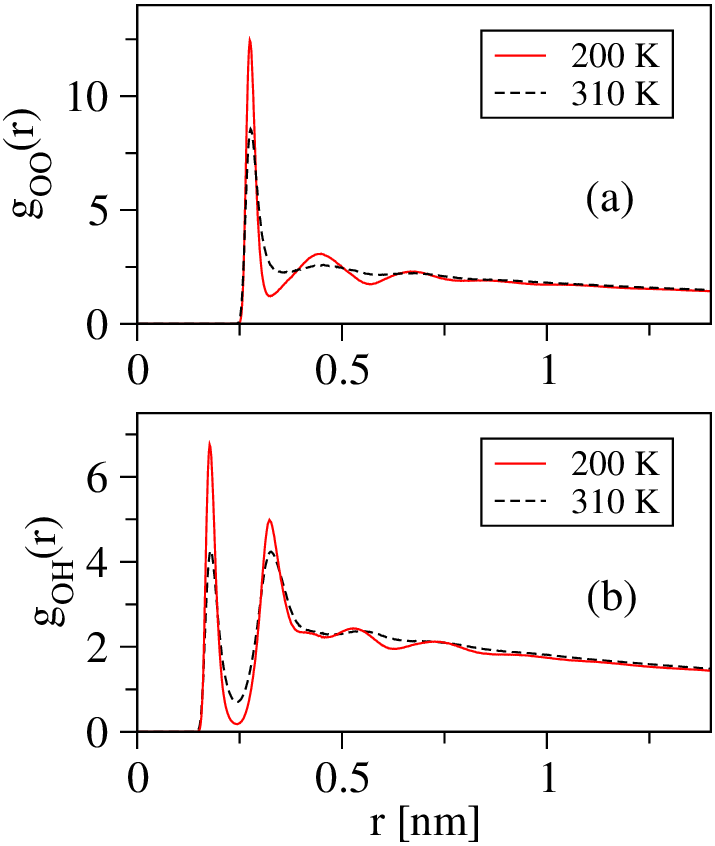}
\caption{Pair distribution functions (a) $g_{OO}$ and (b) $g_{OH}$ for E8 at the
indicated temperatures.} \label{fig1}
\end{center}
\end{figure}

To study the structure of the hydration shells, we show the intermolecular pair distribution functions $g_{OO}(r)$ and $g_{OH}(r)$ for the water oxygens and water hydrogens of E8 in Fig.\ \ref{fig1}. It is evident that the features of $g_{OO}$ and $g_{OH}$ become sharper and an oscillatory behavior develops when the temperature is decreased. The positions of local maxima and minima, which are essentially independent of temperature, are in good agreement with those in previous experimental and computational studies on bulk water, including results for SPC bulk water.\cite{Mark} Specifically, the first maximum of $g_{OO}$ and $g_{OH}$ is located at $r_{OO}\!=\!2.8\mathrm{\,\AA}$ and $r_{OH}\!=\!1.8\mathrm{\,\AA}$, respectively. The first maximum of $g_{OH}$ results from oxygen-hydrogen pairs forming a HB. These findings show that a defined HB network develops at the water-peptide interface upon cooling. In addition, we observe enhanced values $g_{OO}(r)\!>\!1$ and $g_{OH}(r)\!>\!1$ up to $r\!\simeq\!25\mathrm{\,\AA}$, indicating that repulsion of water from the hydrophobic cores of the coiled elastin-like peptides leads to an agglomeration of water between the coils. For none of the models, we find evidence for an onset of crystallization at the studied temperatures.

\begin{figure}
\begin{center}
\includegraphics[width=7cm]{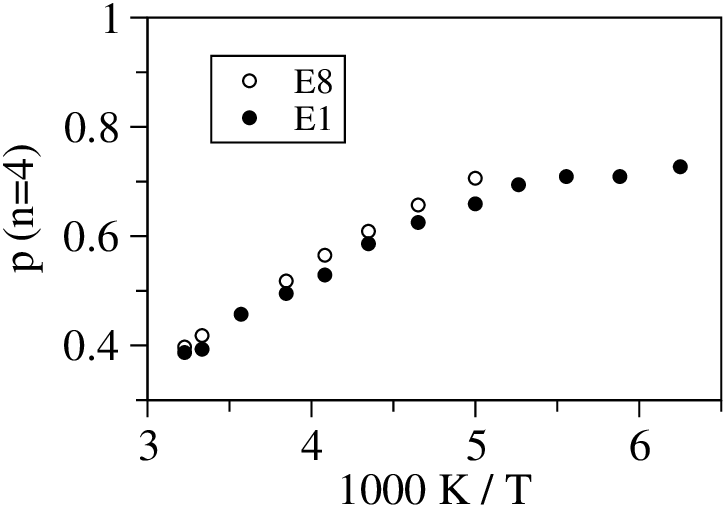}
\caption{Temperature-dependent probability $p$ of finding a water molecule that participates in exactly 4 HB for E1 and E8.} \label{fig2}
\end{center}
\end{figure}

The temperature-dependent structure of the HB networks can be investigated in more detail, when we determine the probability $p$ that a water molecule participates in exactly 4 HB. Exploiting that $g_{OH}(r)$ exhibits a pronounced first minimum, we define that pairs of oxygen and hydrogen atoms form a HB when their interatomic distance is smaller than $2.4\mathrm{\,\AA}$. In doing so, we include water-water and water-peptide HB. In Fig.\ \ref{fig2}, we see that $p$ continuously increases upon cooling until it levels off at about $\mathrm{190\,K}$, at least for E1. These findings suggest that the HB network adopts energetically more favorable configurations upon cooling and, eventually, it finds one of the best possible configurations under the given constraints provided by the water-peptide interface. The absence of a discontinuity implies that a first order phase transition, e.g., between LDL and HDL waters, does not occur at ambient pressure. Due to limited computer power, it was not possible to equilibrate the structure of E8 below $\mathrm{200\,K}$ so as to ascertain whether $p$ saturates for this model as well. 

\subsection{Dynamics of hydration waters}

\begin{figure}
\begin{center}
\includegraphics[width=7cm]{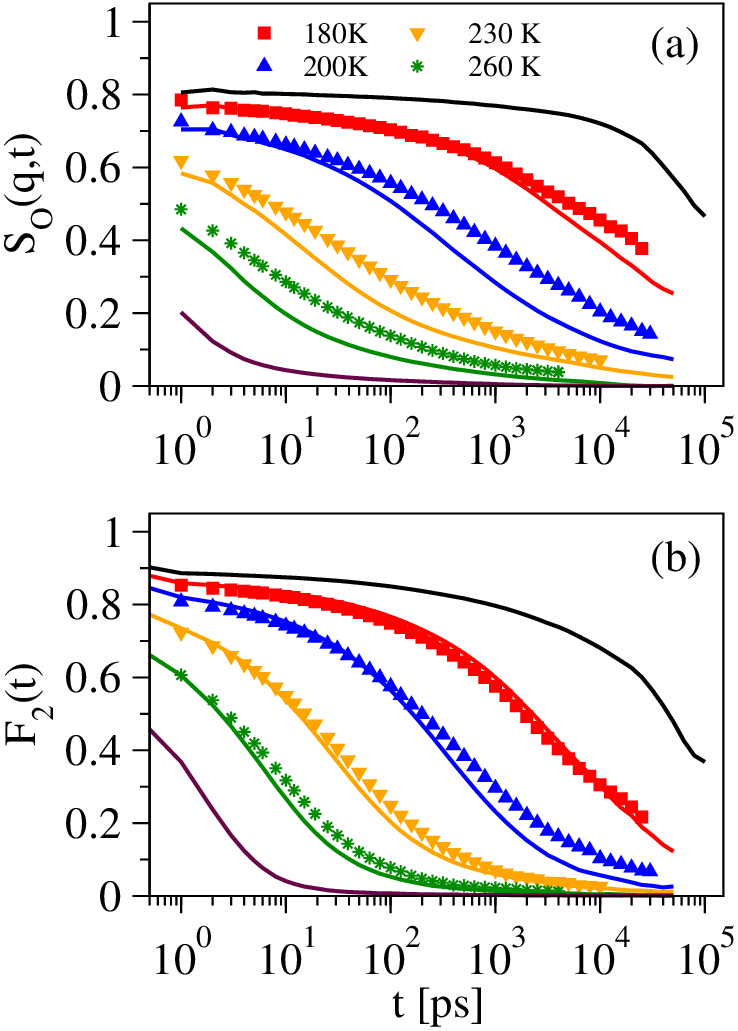}
\caption{(a) Incoherent intermediate scattering functions $S_O(q,t)$ ($q\!=\!2\pi/\mathrm{2.8\,\AA}$) for the oxygen atoms of water in C5 and E1. (b) Orientational correlation functions for the O--H bonds of water in C5 and E1. In both panels, the results for C5 and E1 are shown as symbols and lines, respectively. For C5, the temperatures are indicated. For E1, the temperatures are 310, 260, 230, 200, 180, and $\mathrm{160\,K}$ (from left to right)} \label{fig3}
\end{center}
\end{figure}

To investigate the temperature-dependent dynamical behaviors of the hydration waters, we exploit that the incoherent intermediate scattering function
\begin{equation}\label{ISF}
S(q,t)=\langle\,\cos\{\mathbf{q}\cdot[\,\mathbf{r}(t_0+t)-\mathbf{r}(t_0)]\}\rangle
\end{equation} 
provides us with information about translational motion, while the orientational autocorrelation function 
\begin{equation}\label{F2}
F_2(t)=\frac{1}{2}\,\langle\,3[\,\mathbf{e}(t_0+t)\cdot\mathbf{e}(t_0)]^2-1\rangle
\end{equation}
yields insights into rotational motion. Specifically, the scattering function $S$ depends on the atomic  translational displacements $[\,\mathbf{r}(t_0+t)-\mathbf{r}(t_0)]$ during the time interval $t$, where the absolute value of the scattering vector $q\!=\!|\mathbf{q}|$ determines the length scale on which dynamics is probed. We calculate the scattering functions of the water oxygen atoms, $S_O$, and the water hydrogen atoms, $S_H$, for values of $q$ corresponding to the respective intermolecular interatomic distances. However, we determined that our conclusions do not depend on these specific choices of $q$. The orientational correlation function $F_2$ depends on the angular displacements during the time interval $t$, more precisely, on the value of $|\mathbf{e}(t_0+t)\cdot\mathbf{e}(t_0)|$. Here, the unit vector $\mathbf{e}(t_0)$ describes the orientation of an O--H bond at a time $t_0$. In recent $^2$H NMR work,\cite{Vogel_PRL_08} $F_2$ was determined experimentally for the O--D bonds of heavy water in the hydration shells of elastin and collagen. Throughout this paper, the brackets $\langle\dots\rangle$ denote the average over various time origins $t_0$ and over all atoms or bonds belonging to the considered atomic or bond species.

In Fig.\ \ref{fig3}, we compare temperature-dependent data $S_O$ and $F_2$ for E1 and C5. It is evident that both these functions exhibit strongly nonexpontial decays, which shift to longer times upon cooling. Despite some quantitative differences, $S_O$ and $F_2$ decrease on similar time scales and the behaviors of water near elastin-like and collagen-like peptides are comparable. Consistently, the water dynamics of hydrated elastin and collagen samples were found to be similar in our previous $^2$H NMR study.\cite{Vogel_PRL_08} For the higher of the studied temperatures, the observation of complete decays indicates that the $\alpha$ relaxation of hydration water is probed. Specifically, the results for $S_O$ show that all water molecules move at least one intermolecular distance on a time scale of a few nanoseconds at $\mathrm{260\,K}$. Nonexponential $\alpha$ relaxation is a characteristic feature of supercooled liquids approaching their glass transitions. While a stretched exponential function well describes the time dependence of the $\alpha$ relaxation in these materials, such function does not yield a reasonable interpolation in our case. We will return to this point in Sec.\ \ref{interface}. 

\begin{figure}
\begin{center}
\includegraphics[width=7cm]{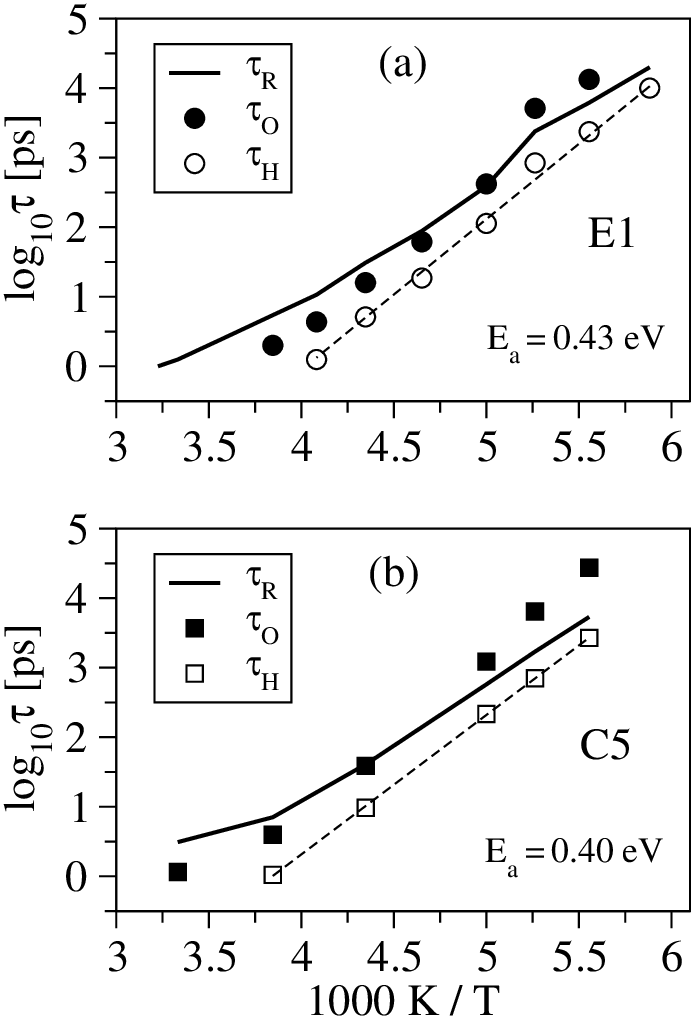}
\caption{Correlation times of water dynamics in (a) E1 and (b) C5. In both panels, we compare translational correlation times $\tau_O$ and $\tau_H$ for the oxygen atoms from $S_O(q,t)$ ($q\!=\!2\pi/\mathrm{2.8\,\AA}$) and for the hydrogen atoms from $S_H(q,t)$ ($q\!=\!2\pi/\mathrm{2.4\,\AA}$), respectively, with rotational correlation times $\tau_R$ from $F_2(t)$. All shown values are 1/e-decay times, e.g., $F_2(\tau)\!=\!e^{-1}$. The dashed lines are Arrhenius fits to $\tau_H$.} \label{fig4}
\end{center}
\end{figure}

To study the slowdown of water dynamics upon cooling, we determine the translational correlation times $\tau_O$ and $\tau_H$ and the rotational correlation times $\tau_R$ from the corresponding incoherent scattering and orientational correlation functions. Figure \ref{fig4} shows the temperature dependence of these time constants for E1 and C5. We see that $\tau_H$ follows an Arrhenius law with comparable activation energies $E_a\!=\!0.40\!-\!0.43\mathrm{\,eV}$ for both models. Such activation energies, which are typical of breaking of 2 HB, were reported for water dynamics in various types of confinement at sufficiently low temperatures.\cite{Swenson_PRL_06} The temperature dependence is weaker for $\tau_O$ and $\tau_R$ at high temperatures. Thus, there are deviations from an Arrhenius law, which are typical of fragile supercooled liquids like weakly supercooled bulk water.\cite{Angell_SCI_95} Here, the different temperature dependences of $\tau_O$ and $\tau_H$ imply that diverse mechanisms for water dynamics exist, including anisotropic motions. While $\tau_O$ reflects the translational motion of the water molecule, the value of $\tau_H$ depends on both translational and rotational motions. For example, $\pi$ flips about the twofold molecular symmetry axis would affect $\tau_H$, but not $\tau_O$, see Sec.\ \ref{mech}.      

\begin{figure}
\begin{center}
\includegraphics[width=7cm]{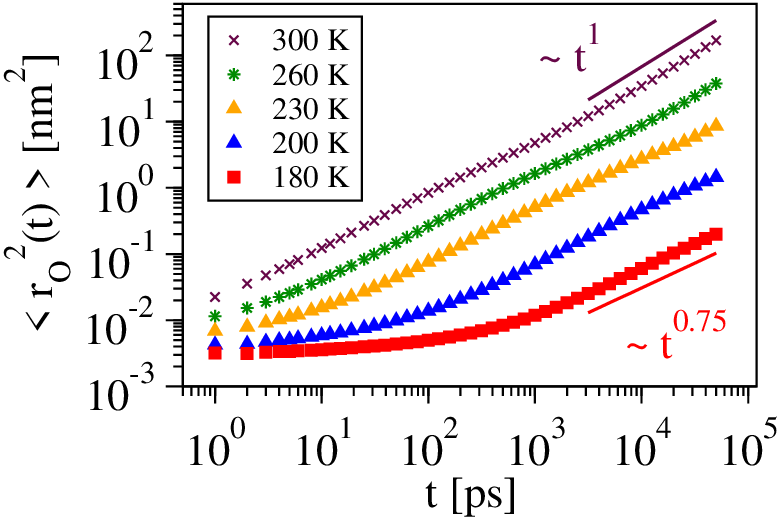}
\caption{MSD for the oxygen atoms of water in E1 at various temperatures. The lines are power laws $\langle r^2(t)\rangle \propto t^1$ and $\langle r^2(t)\rangle \propto t^{0.75}$, respectively.} \label{fig5}
\end{center}
\end{figure}

Next, we discuss the atomic mean-square displacements (MSD) 
\begin{equation}\label{msd}
r^2(t)=\langle\,[\mathbf{r}(t_0+t)-\mathbf{r}(t_0)]^2\rangle
\end{equation}
Figure \ref{fig5} shows the temperature-dependent MSD of the water oxygens in E1, which is a measure for the center-of-mass motion of water. In addition to a ballistic regime, which is expected for short times outside the studied time window, three regimes exist for the translational motion of water. At short times and low temperatures, we observe a plateau regime, indicating that the water molecules are trapped in cages formed by neighboring water and peptide molecules. This cage effect is typical of supercooled liquids approaching their glass transitions. At longer times, the water molecules start to escape from these cages and, hence, the MSD increases. Subsequent to the plateau, we find a regime of sublinear diffusion, $r^2\!\propto\!t^{0.75}$, reflecting the effect that the presence of large peptide chains prevents water from free diffusion. In harmony with these results, a recent study on the hydration water of RNase A found that a plateau of the MSD develops between a regime of ballistic motion and a regime of sublinear diffusion when the temperature is decreased.\cite{Tarek_EBPJ_08} Finally, at high temperatures and long times, there is a crossover to linear diffusion, $r^2\!\propto\!t$, indicating that the spatial confinement due to the peptide becomes unimportant at sufficiently large length scales. These three regimes are also observed for water diffusion in C5 and E8 and, hence, their existence is not an artifact of the small size of model E1. For E1, a confinement effect results since the water molecules do not only interact with the single peptide chain in the simulation box, but, because of the use of periodic boundary conditions, also with all its periodic images. The models C5 and E8 contain several peptide molecules so that there are also confinement regions within the simulation box.

When the temperature is decreased, the plateau regime extends, indicating that the water molecules become trapped for longer and longer times in their local cages. Despite this strong delay of water transport, we do not find evidence that the diffusion of the hydration water ceases, but it rather exits the time window of the simulation in a continuous manner. The relations between water diffusion and protein dynamics will be discussed in future work. While a self-diffusion coefficient of $D\!=\!5.6 \cdot 10^{-10}\mathrm{\,m^2/s}$ results from the linear regime of the MSD for the hydration water of E1 at $300\mathrm{\,K}$, a value of $D\!\approx\!4.2 \cdot 10^{-9}\mathrm{\,m^2/s}$ was reported for SPC bulk water at $298\mathrm{\,K}$.\cite{Mark,Spoel} Thus, the presence of the elastin-like peptide leads to a slowdown of water diffusion by a factor of $7\!-\!8$ at ambient temperatures. 

\subsection{Mechanisms for the dynamics of hydration waters}\label{mech}

\begin{figure}
\begin{center}
\includegraphics[width=7cm]{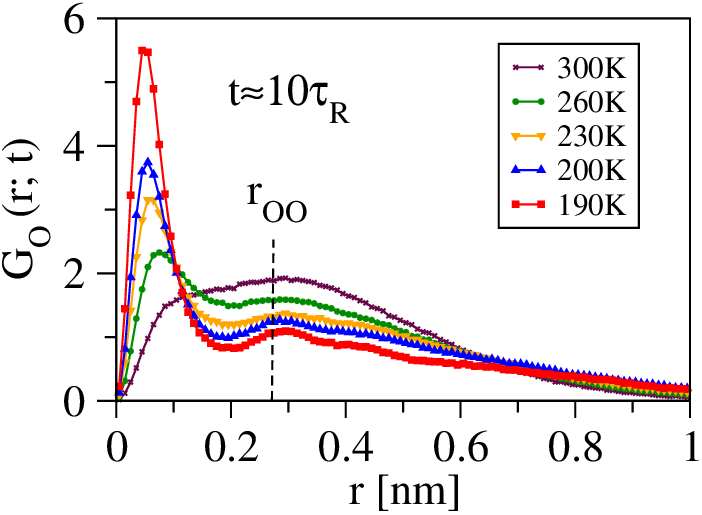}
\caption{Self part of the van Hove correlation function $G_O(r;t)$ for the oxygen atoms
of water in E1. For all studied temperatures, we use comparable time intervals $t\approx\!10
\tau_R$. The dashed line marks the interatomic oxygen-oxygen distance $r_{OO}$.} \label{fig6}
\end{center}
\end{figure}

The mechanisms for the translational motion of water can be investigated on the basis of the van Hove self-correlation function of the water oxygens
\begin{equation}\label{van}
G_O(r,t)=\langle\,\delta[r-|\,\mathbf{r}(t_0+t)-\mathbf{r}(t_0)|]\rangle
\end{equation}
It measures the probability that a water oxygen moves a distance $r$ in a time interval $t$. In simple cases, the van Hove correlation function is a Gaussian. Figure \ref{fig6} shows $G_O(r;t)$ of E1 for various temperatures and comparable time intervals, which correspond to late stages of the $\alpha$ relaxation, see Fig.\ \ref{fig3}. It is evident that the Gaussian approximation is not obeyed at any of the studied temperatures. Upon cooling, $G_O(r;t)$ develops a multi-peak structure, where the positions of the secondary maxima coincide with the positions of the peaks of $g_{OO}(r)$, see Fig.\ \ref{fig1}. In particular, a well resolved secondary maximum exists at $r\!\approx\!r_{OO}$. These observations show that, when the temperature is decreased, a HB network with defined sites develops and the energy barriers between these sites start to govern water dynamics, leading to a jump motion of the water molecules. Here, the term jump means that long periods of vibrational motion about the sites are interrupted by short periods necessary to cross the saddles that separate the sites in the HB network.   

\begin{figure}
\begin{center}
\includegraphics[width=7cm]{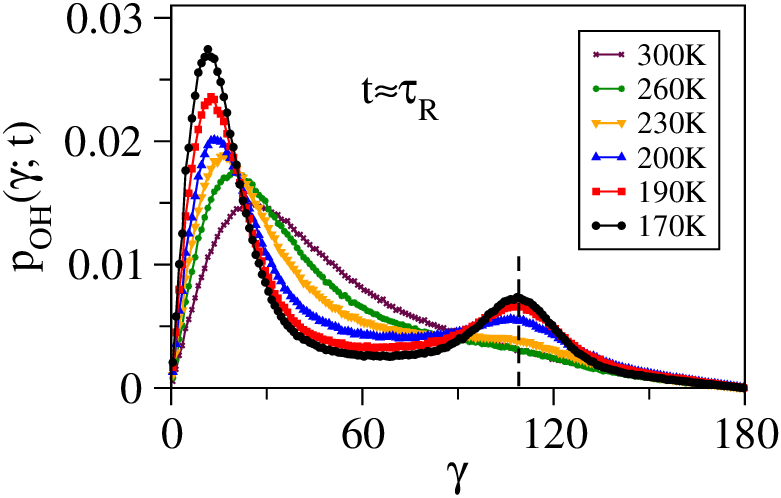}
\caption{Probability distributions $p_{OH}(\gamma;t)$ of finding an angular displacement $\gamma$ during a time interval $t$ for the O--H bonds of water in E1. For all studied temperatures, we use comparable time intervals $t\approx\!\tau_R$. The dashed line marks the tetrahedral angle.} \label{fig7}
\end{center}
\end{figure}

Valuable information about the mechanisms for the rotational motion of water can be obtained when we determine the probability distribution $p(\gamma;t)$ of finding an angular displacement $\gamma$ in a time interval $t$. In other words, $p(\gamma;t)$ describes the probability that the orientations of a given bond at the beginning and at the end of the time interval differ by an angle $\gamma$. First, we study $p_{OH}(\gamma;t)$ for the O--H bonds of water. In Fig.\ \ref{fig7}, we display $p_{OH}$ of E1 for various temperatures and $t\!\approx\!\tau_R$. Upon cooling, there is a continuous crossover from a one-peak to a two-peak signature, indicating that the rotational dynamics of water evolves from small-angle to large-angle motion. The position of the secondary maximum shows that the O--H bonds tend to jump about angles close to the tetrahedral angle at low temperatures.       

\begin{figure}
\begin{center}
\includegraphics[width=7cm]{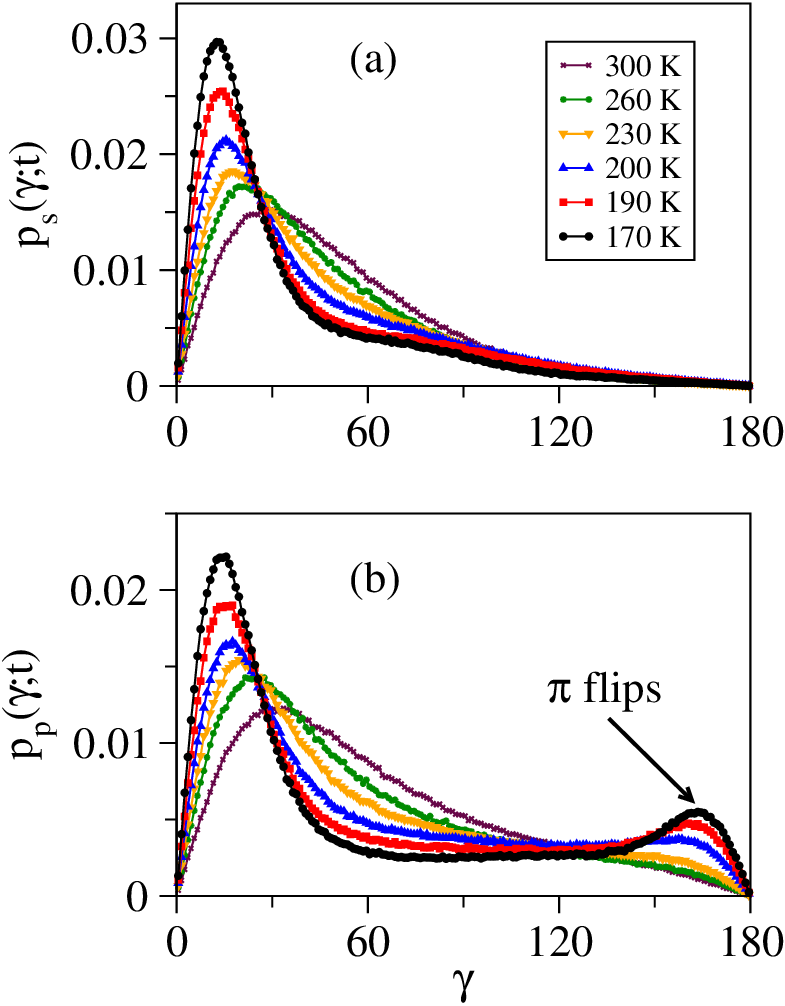}
\caption{Comparison of the probability distributions $p_{s}(\gamma;t)$ and
$p_{p}(\gamma;t)$ providing information about the temperature-dependent anisotropy of
water reorientation in E1. While $p_{s}$ describes the angular displacements of the
two-fold symmetry axes of the water molecules, $p_{p}$ characterizes the angular
displacements of a vector perpendicular to the plane defined by the three atoms of the
respective molecule. For both probability distributions, we use comparable time intervals
$t\approx\!\tau_R$ at the indicated temperatures.} \label{fig8}
\end{center}
\end{figure}

To obtain further insights into the nature of the rotational motion, we investigate the reorientation of two orthogonal vectors that are linked to the molecular frame of water. Specifically, we calculate $p_{s}(\gamma;t)$ and $p_{p}(\gamma;t)$, which describe the rotational motions of the two-fold symmetry axis of the water molecule and of the normal vector of the plane formed by the atoms of the water molecule, respectively. In Fig.\ \ref{fig8}, we show both quantities for E1 at various temperatures using $t\!\approx\!\tau_R$. We see that a secondary maximum exists for $p_p$, but not for $p_s$. The latter finding means that the reorientation of the two-fold axis is small during the jump motion of most molecules, i.e., this axis essentially coincides with the axis of rotation. The position of the secondary maximum of $p_p$ indicates that the normal vector performs jumps by about $180^{\circ}$. Together, these results show that distorted $\pi$ flips about the two-fold symmetry axes are an important aspect of water dynamics at sufficiently low temperatures, at least in our models.         

\subsection{Effects resulting from the peptide interfaces}\label{interface}

\begin{figure}
\begin{center}
\includegraphics[width=7cm]{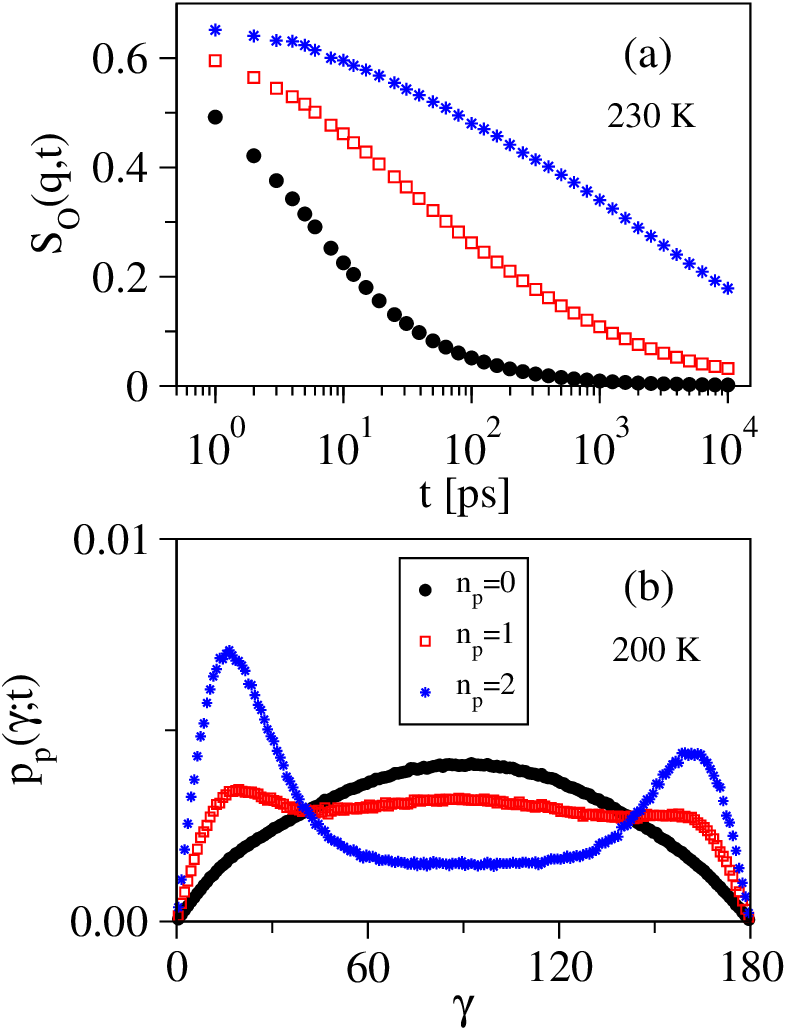}
\caption{Separate analyses of water dynamics in E8 for water molecules forming different numbers $n_p$ of hydrogen bonds with proteins: (a) Incoherent intermediate scattering functions $S_O(q,t)$ ($q\!=\!2\pi/\mathrm{2.8\,\AA}$) for the oxygen atoms of water at $\mathrm{230\,K}$ and (b) probability distribution $p_{p}(\gamma,t)$ for $t\!=\!10\mathrm{\,ns}$ at $\mathrm{200\,K}$.} \label{fig9}
\end{center}
\end{figure}

Finally, we study whether water dynamics changes in the vicinity of elastin-like peptides. For this purpose, we distinguish between water molecules forming different numbers $n_p$ of HB to the peptides in our analyses. In Fig.\ \ref{fig9}, we present the outcome of such separate analyses for the example of E8. In panel (a), we compare the incoherent scattering functions $S_O(q,t)$ of the different water species at $230\mathrm{\,K}$. We see that the decay resulting from water molecules that form 2 HB with peptides is about three orders of magnitude slower than that resulting from water molecules forming no HB with peptides. Thus, as a consequence of HB formation, the immediate vicinity of a peptide leads to a strong slowdown of the water dynamics. These pronounced dynamical heterogeneities result in a pronounced non-exponentiality of the translational and rotational correlation functions for the whole ensemble of water molecules, see Fig.\ \ref{fig3}. In particular, they result in deviations from a stretched-exponential behavior, which is usually found for supercooled liquids. Specifically, a long-time tail results from water molecules that are more tightly bound to a peptide through two HB. Comparable results are obtained for E1 and C5.

In panel (b), the heterogeneity of water dynamics is demonstrated for the rotational motion on the basis of $p_p(\gamma;t)$ at $200\mathrm{\,K}$. For water molecules with $n_p\!=\!0$, we see that the distribution is close to the final state $p(\gamma,t\!\rightarrow\!\infty)\!\propto\!\sin \gamma$, indicating that the normal vectors have been isotropically redistributed on the unit sphere during the used time interval $t\!=\!10\mathrm{\,ns}$. By contrast, for water molecules with $n_p\!=\!2$, the intensity is cumulated in two peaks at small and large angles, respectively, indicating that most molecules in the immediate vicinity of a peptide have only performed distorted $\pi$ flips during this time interval.         

\section{Discussion and Conclusion}

We have investigated the temperature-dependent water dynamics in the hydration shells of elastin-like and collagen-like peptides using MD simulations. Analysis of the simulation data shows that the water dynamics hardly depend on the type of the peptide. For all studied models, water dynamics is strongly slowed down in the immediate vicinity of the peptides, in particular at low temperatures, because of formation of HB between water and peptide molecules, leading to pronounced nonexponential correlation functions of water motion. The mechanisms for both translational and rotational water dynamics change from small-step (diffusive) motion to large-step (jump) motion upon cooling. Concerning the rotational motion, we have found that distorted $\pi$ flips of the water molecules about their twofold symmetry axes become important when the temperature is decreased, at least for molecules at the water-peptide interfaces, i.e., a water-specific $\beta$ process evolves. While this $\beta$ process of water is associated with large-angle jumps, the JG $\beta$ process results from small-angle motion,\cite{Vogel_JCP_01} suggesting that both secondary relaxations are not directly related. In harmony with the present results, the hydration waters of lysozyme were found to perform large-angle jumps below ambient temperatures.\cite{Jana_JPCB_08}

Furthermore, we have observed complex water self diffusion, which is characterized by four time regimes at sufficiently low temperatures: a ballistic regime, a plateau regime, a regime of sublinear diffusion, and a regime of linear diffusion. When the temperature is decreased, the plateau regime extends, indicating that the water molecules become trapped for longer and longer times in cages formed by their neighbors. A regime of sublinear diffusion regime was also observed for the hydration waters of RNase A.\cite{Tarek_EBPJ_08} It results because the larger peptide chains form some confinement for the water molecules and, thus, hamper free diffusion, which is only observed at sufficiently large time and length scales. The existence of pronounced sublinear diffusion means that determination of reliable diffusion coefficients from MD simulations requires relatively long equilibration and production runs.     

For the temperature dependence of various correlation times, e.g., of the rotational correlation time, we have found a weak crossover from a non-Arrhenius behavior at high temperatures to an Arrhenius behavior at low temperatures, which occurs at $\tau\!\approx\!1\mathrm{\,ns}$. A weak crossover was also reported for the water self-diffusion coefficient in simulation work using the TIP5P model of water together with the GROMOS model of lysozyme.\cite{Kumar_PRL_06} However, the present findings are contrary to computational results for TIP4P-Ew water in the hydration shell of OPLS-AA lysozyme.\cite{Lagi_JPCB_08} Despite comparable correlation times of a few nanoseconds at the crossover in the present and previous studies, a sharp kink toward an Arrhenius law characterized by a low activation energy of $E_a\!\approx\!0.15\mathrm{\,eV}$ was observed in the literature, while we find a weak crossover to a significantly higher value of $E_a\!\approx\!0.43\mathrm{\,eV}$ for both translational and rotational correlation times. The former value of the activation energy is consistent with results from neutron scattering work,\cite{Chen_PNAS_06} whereas the latter is in agreement with activation energies determined in DS and NMR studies.\cite{Cerveny_PRL_04,Vogel_PRL_08} The origin of these deviations is not yet understood. To clarify this point, it is desirable to compare simulation results for various water and protein force fields in future work.   

In addition, we have demonstrated that an increase of the order of the HB network structure accompanies the change of the mechanism for the water dynamics and the weak crossover from fragile behavior to strong behavior. Specifically, the percentage of water molecules exhibiting an ideal number of exactly 4 HB continuously increases upon cooling until it saturates at temperatures $190\!-\!210\mathrm{\,K}$, at which the temperature-dependent correlation times show the weak crossover. This agreement suggests an intimate relation between structure and dynamics of the hydration water. When the temperature is decreased the energetic penalty a water molecule experiences when being in an imperfect HB environment becomes more and more important, leading to constraints for the molecular positions and orientations. As a consequence, the mechanism for water dynamics changes from diffusive motion to jump motion and a local $\beta$ relaxation sets in prior to the global $\alpha$ relaxation. These conclusions are in harmony with results of recent calculations for a cell model of water.\cite{Kumar_PRL_08} There, it was argued that, when the temperature is decreased, the number of HB increases and the orientational disorder decreases, resulting in a rise of the activation energy and, hence, to non-Arrhenius behavior. When further cooling, the rate of change of orientational disorder reaches a maximum and, then, it rapidly drops to zero, leading to temperature independent both disorder and activation energy at low temperatures. 

As compared to experimental results, the dynamics of SPC water is known to be too fast.\cite{Spoel,Mark} Likewise, the water dynamics in our models is faster than that in hydrated elastin and hydrated collagen. Specifically, $\tau_R\!=\!1\mathrm{\,ns}$ at $200\mathrm{\,K}$ in the present study and at $250\mathrm{\,K}$ in our previous $^2$H NMR spin-lattice relaxation analysis.\cite{Vogel_PRL_08} This discrepancy hampers straightforward comparison of computational and experimental correlation times and crossover temperatures. Moreover, it is not clear whether results for the motional mechanism from MD simulations and $^2$H NMR stimulated-echo experiments may be compared since water dynamics in the nanosecond regime and in the millisecond regime are studied, respectively. A common activation energy of $E_a\!\approx\!0.45\mathrm{\,eV}$ suggests that both techniques probe the same low-temperature dynamical process of protein hydration water despite the different time windows. In harmony with the present results, we found in our previous NMR study that the mechanism for water dynamics changes from an isotropic motion to an anisotropic large-angle jump motion upon cooling.\cite{Vogel_PRL_08} While exact $\pi$ flips were ruled out as motional mechanism in the low-temperature Arrhenius regime, inexact $\pi$ flips may be consistent with our $^2$H NMR data. Moreover, exact $\pi$ flips cannot be observed in DS, in contrast to experimental findings,\cite{Bergman_NAT_00,Cerveny_PRL_04,Swenson_PRL_06} but the small-angle wobbling motion of the twofold symmetry axes of the water molecules associated with distorted $\pi$ flips would be probed by the latter method. Further experimental and computational studies are required to clarify these points. In particular, it is important to determine whether the existence of distorted $\pi$ flips is common to various established water and protein force fields.  

\section{Acknowledgement}
Funding of the DFG through grants VO 905/3-1 and VO 905/3-2 is gratefully acknowledged.


\begin{thebibliography}{99}
\bibitem{Hyp} The letter O is used as single-letter code of hydroxyproline.
\end{thebibliography}
\end{document}